\documentclass{sig-alternate_WEDBT2016}

\usepackage[
  pass,
]{geometry}

\usepackage{pifont}
\usepackage{marvosym}
 \usepackage{times}%
\usepackage[T1]{fontenc}
 \usepackage[table]{xcolor}

 \usepackage[caption=false]{subfig}
\usepackage{color}
\usepackage{multirow}
 \usepackage{graphicx}
\usepackage[linesnumbered, vlined, ruled,algo2e]{algorithm2e} 
\usepackage{booktabs}
\usepackage{amssymb}
\usepackage{amsfonts}
\usepackage{amsmath}
\usepackage{mathtools}

\usepackage{amsthm}
\usepackage{soul}
\usepackage{etex}
 \usepackage{relsize}
\usepackage{epstopdf}
\usepackage{multirow}
\usepackage{url}
\usepackage{tablefootnote}
 
 \usepackage{hyperref}  
\hypersetup{ 
    colorlinks,%
    citecolor=black,%
    filecolor=black,%
    linkcolor=black,%
    urlcolor=black,%
pdftitle={ Exploration and Visualization in the Web of Big   Linked Data:   A Survey of the State of the Art and Challenges Ahead},
pdfauthor={Nikos Bikakis, Timos Sellis},
pdfsubject={Exploration and Visualization in the Web of Big   Linked Data:   A Survey of the State of the Art and Challenges Ahead},
pdfproducer={pdfbikTeX-1.1},
pdfcreator={bikTeX},
pdfkeywords={survey, visualization, big data, exploration, state of the art, challenges, visual analytics, study, data exploration, big data challenges, 
information visualization, state-of-the-art, large databases, scalability, semantic web, linked data, web of data, web of linked data, open data, study,  survey, large data visualization, navigation, RDF, visualizing very large datasets,  incremental visual exploration, progressive visualization, incremental visualization, multiresolution visualization, multiscale visualization, multilevel visualization, hierarchical visualization, adaptive visualization, data summarization, data abstraction, binning, binning aggregation, visualization recommendations, tree reconstruction, dynamic tree, Steerable Exploration, visualization preferences, scalable visualizations, visual aggregation, personalized exploration, semantic web, hierarchical navigation, personalized navigation, pay as you go visualization,  visualization, linked data visualization, web of data, SPARQL, hierarchical aggregation, OWL, scalable visual analytics,  hierarchical exploration, Steerable visualization, scalable exploration, multiresolution exploration, multiscale exploration, multilevel exploration, incremental exploration, personalized visual exploration, multiresolution navigation, multilevel navigation, incremental navigation, multiscale navigation, personalized visualization, visualization tree, personalized visual analytics, exploration user preferences, multiresolution visual analytics, hierarchical charts, RDF Facets, HETree, Faceted search, multilevel temporal visualization, LOD, rdf:SynopsViz, hierarchical timeline, linked open data, multilevel timeline, hierarchical histogram, tree visualization structure, numeric exploration, LD, multilevel charts, RDF Charts, Data exploration, RDF Statistics, temporal exploration, synopsviz, scalable navigation, personalization, , multilevel numeric visualization, multi-scale, multi-level, pay-as-you-go visualization, multi-resolution}
}

\newcommand{\eat}[1]{}

\newcommand{\stitle}[1]{\vspace{0.2cm}\noindent\textbf{#1}}

\definecolor{dark-gray}{gray}{0.2}

\renewcommand{\qed} {\hfill$\blacksquare$}

\newcommand*\yes{\ding{51}}
\newcommand*\no{}

\newcommand{\tline}  {\specialrule{0.8 pt}{0pt}{1pt}}		 
\newcommand{\bline}  {\specialrule{0.8 pt}{1pt}{0pt}}		
\newcommand{\dlineB}  {\specialrule{0.8 pt}{1pt}{2pt} \specialrule{0.8 pt}{0pt}{0pt}}		

\newcolumntype{'}{!{\vrule width 0.8pt}}

\SetKw{Logand}{and}	
\SetKw{Logor}{or}
\SetKw{Lognot}{not}	

\SetKw{mybreak}{break}	
\SetKw{mycontinue}{continue}	

\SetKwInput{KwVar}{Variables}
\SetKwInput{KwPar}{Parameters}

\SetKwData{myEOF}{EOF}	
\SetKwData{myfalse}{false}
\SetKwData{mytrue}{true}
\SetKwData{mynull}{null}
\SetKwData{myempty}{empty}
\SetKwBlock{myblock}{beginnikos}{end}

 \SetAlgoInsideSkip{} 
\SetAlgoSkip{} 
\SetAlgoCaptionSeparator{.} 

\newcommand{\mycomment} [1] 	{\tiny{\textcolor{dark-gray}{\myFontP{{#1}}}}} 
\SetKwComment{Comment}{\mycomment{/\!\!/}}{}
 \DontPrintSemicolon
\SetAlgoCaptionLayout{small}
\SetProcNameSty{small}
\SetProcArgSty{small}

\begin{document}
 



\conferenceinfo{LWDM `16}{March 15, 2016, Bordeaux, France}



\title{  Exploration and Visualization in the Web of Big  
Linked Data: 
  A Survey of the State of the Art 
\titlenote{This paper appears in 6th International Workshop on Linked Web Data Management (LWDM 2016).}
}

\numberofauthors{2} 

\author{
\alignauthor Nikos Bikakis\\
\affaddr{NTU Athens \& \\ATHENA R.C.} \\
\affaddr{Greece}\\
\alignauthor Timos Sellis\\
\affaddr{Swinburne Univ.\\\ of Technology}\\
\affaddr{Australia}\\
}

\maketitle

\begin{abstract}
Data exploration and visualization systems are of great importance in the  \textit{Big Data}  era. 
Exploring and visualizing very large datasets has become a major research challenge, of which {scalability} is a vital requirement.  
In this survey,  we describe the major prerequisites  and challenges  that should be addressed by the modern exploration and visualization systems. 
Considering these challenges, we present how   \textit{state-of-the-art} approaches from the \textit{Database} and \textit{Information Visualization} communities 
attempt to handle them. 
Finally, we survey the systems developed by \textit{Semantic Web} community in the context of the \textit{Web of Linked Data},  and   discuss to which extent these satisfy the contemporary requirements.
\vspace{-1mm}
\end{abstract}


\keywords{Visual analytics, big data challenges, data exploration, large databases, visual exploration, semantic web, visualization tools, scalability}


\section{Introduction}
\label{sec:intro}

The purpose of \textit{data exploration} and \textit{visualization} is to offer ways for information perception and manipulation, as well as knowledge extraction and inference \cite{IdreosPC15,HeerS12}. 
Data visualization\footnote{Throughout this paper we use the term "visualization" referring to visual data exploration.\vspace{-1.2cm}} provides  users with an intuitive means to explore the content of the data, identify interesting patterns, infer correlations and causalities, and supports sense-making  activities.
Data exploration and visualization systems are of great importance in the  \textit{Big Data}  era, 
in which  the volume and heterogeneity of available information make it   difficult for humans to manually explore and analyse data. 

Most traditional systems cannot handle the  {large size} of many contemporary  datasets. Exploring and visualizing large datasets has become a major research challenge \cite{bp16,Shneiderman08,HeerK12b, MortonBGM14,WuBM14,GGL15}.
Therefore, modern systems have to take into account \textit{scalability}, as a vital requirement.  
Dealing with scalability, modern systems have to address numerous issues related to  storage, access, rendering/presentation, interaction, etc. 

In the \textit{Web of Data} (WoD) context,  following the abundance of \textit{Linked  Data},  
several recent efforts have offered tools and techniques for  exploration and  visualization in many different domains \cite{DR11}.
However, most of these approaches fail to take into account issues related to performance and scalability. 

In this work, we describe the major requirement and challenges that should be addressed by the modern exploration and visualization systems. 
Additionally, we refer to state-of-the-art approaches from the Database and Information Visualization communities, which attempt to handle some of these challenges. 
Further, we describe the systems that have been developed in the context 
of WoD, and   discuss to which extent they  satisfy the contemporary requirements. 





 \section{Challenges}
 \label{sec:chall}

Most traditional exploration and visualization systems operate in an \textit{offline} way, limited to accessing static sets of preprocessed data. 
Additionally, they restrict themselves to dealing with \textit{small} dataset sizes, which can be easily handled and explored with conventional data management and (visual) explorations techniques. 

On the other hand, nowadays, the Big Data era has realized the availability of the great number and variety of \textit{very large} datasets that are \textit{dynamic} in nature.
For example, most data sources offer query or API endpoints   for
online access and updates;  in other cases (e.g.,
scientific databases), new data is constantly arrived (e.g., on a
daily/hour basis). 
Beyond these, modern systems should operate on exploratory
context. In an \textit{exploration scenario}, it common that
users are interesting in finding something interesting and useful
without previously know what exactly are searching for, until
the time they identify it. In this case, users perform a sequence
of operations (e.g., queries), in which the result of each
operation determine the formulation of the next operation.
Finally,  an increasingly  large number of \textit{diverse users} (i.e., different preferences, skills, etc.)  explore and analyse data  in a plethora of  \textit{different scenarios}.

Therefore, some of the major challenges  that should be dealt with by  modern systems,  are posed by the: 
(1) Large size and the dynamic nature of  data in conjunction with  
the exploration-driven setting; and (2)  Variety of tasks and users.

%
%

\stitle{Large \& Dynamic Data in Exploration-driven Setting.}
One of the major challenges in exploration and visualization is
related to the  \textit{size} that characterizes   most contemporary datasets. 
A second challenge is related to the   availability of  query  and API endpoints  for online data access and retrieval, as well as the cases where  that data is received in a stream fashion. 
The later pose  the challenge of
handling large sets of data in a \textit{dynamic} setting, and as a result,
a preprocessing phase (e.g., traditional indexing) is prevented.


In this respect, modern visualization and exploration  systems must be able to \textit{efficiently} and \textit{effectively} handle \textit{billion objects dynamic datasets} throughout an
\textit{exploratory scenario}. 
Therefore, scalable  and  {efficient} data structures and algorithms have to be developed. 
Crucial issues related to storage,   access, management, presentation,  interaction (e.g., pan, zoom, search, drill-down), etc. over large dynamic datasets have to be handled. 
Scalability has become a major challenge for the modern   systems. 
Beyond these,  systems have to efficiently operate on machines with limited computational and memory resources (e.g.,  laptops). 



In a "conventional" setting (e.g., explore a small fragment of a preprocessed dataset), most of the aforementioned issues can be handled by 
the traditional  systems that provide database exploration and analysis, 
such as  
Tableau\footnote{\href{tableau.com}{tableau.com}}   (previously know as Polaris \cite{StolteH00}),
DEVise \cite{LivnyRBCDLMW97a},
Spotfire~ \cite{Ahlberg96},  
VisDB \cite{KeimK94},
Lumira\footnote{\href{sap-lumira.com}{sap-lumira.com}}, 
QlikView\footnote{\href{clickview.com}{clickview.com}}, 
Datawatch\footnote{\href{datawatch.com}{datawatch.com}}, etc.
However, in a "modern" setting, when  a large part (or the whole)  of a billion objects dynamic  dataset has to be explored,   
the aforementioned traditional database-oriented systems cannot be  adopted. 
 
 
In conjunction with performance issues, modern systems have to 
address challenges related to visual presentation and interaction issues. 
Particularly, systems  should be able to present, as well as, offer ways to "easily" explore large datasets. 
Handling a large number of data objects is a challenging task;  modern systems have to 
"\textit{squeeze a billion records into a million pixels}" \cite{Shneiderman08}. 
Even, in  much smaller datasets,  offering a dataset overview  is  extremely difficult;
in both cases  \textit{information overloading} is a common issue.
As aslo stated in the visual information seeking mantra: "\textit{overview first, zoom and filter, then details on demand}" \cite{Shneiderman96}, 
gaining \textit{overview} is crucial   in the visual exploration scenario. 
Based on the aforementioned, it follows that a basic requirement of the modern systems  
is to develop methods that provide \textit{summaries} and \textit{abstractions} 
over the enormous number of data objects.

In order to tackle both performance and presentations
issues, a large number of  systems adopt \textit{approximation techniques} (a.k.a.\ \textit{data reduction} techniques)
in which  partial results are computed. 
Considering the existing approaches, most of them are  based on: 
(1) sampling and filtering  \cite{FisherPDs12,ParkCM15,AgarwalMPMMS13,ImVM13,BattleSC13}; 
or/and  (2) aggregation (e.g., binning, clustering) \cite{EF10, bsps15,JugelJM15, JugelJHM14a,LiuJH13,hw13,LinsKS13,AbelloHK06,BastianHJ09,RodriguesTPTTF13}. 
In this respect,  some modern database-oriented systems adopt approximation techniques 
using query-based approaches (e.g., query translation, query rewriting) \cite{BattleSC13,JugelJM15,JugelJHM14a}.

In order to improve efficiency several systems adopt \textit{incremental} (a.k.a.\ \textit{progressive}) techniques.
In these techniques the results/visual elements are computed/constructed incrementally based on  user interaction or as time progresses (e.g., \cite{StolperPG14, bsps15}).
Numerous recent systems integrate incremental and approximate techniques, in these approaches, approximate answers are computed incrementally over progressively larger samples of the data \cite{FisherPDs12,AgarwalMPMMS13,ImVM13}.

The dynamic setting prevents modern systems from preprocessed the  data.
Additionally, it is common in exploration scenarios
only a small fragment of data to be accessed by the user. 
In this context,   an adaptive indexing approach \cite{IdreosKM07} is used in \cite{ZoumpatianosIP14}, where the indexes are created incrementally and adaptively throughout
exploration. 
Similarly, in \cite{bsps15} the hierarchy tree is
incrementally constructed based on user's interaction.
Finally, in some approaches,  parallel architectures are adopted; e.g., \cite{EMJ16,KamatJTN14, KalininCZ15,ImVM13}.

To sum up, modern systems should provide scalable techniques that on-the-fly effectively (i.e., in a way that can be easily explored) handle a large number of data objects 
over an exploration scenario, 
using a limited number of resources





%
%

 \vspace{2mm}
\stitle{Variety of  Tasks \& Users.}
The requirement of scalable, on-the-fly  exploration and analysis  must be coupled with the diversity of preferences and requirements posed by different users  and tasks. 

Therefore,  the modern systems should provide the user with the ability to customize the exploration experience based on her preferences and the requirements posed by the examined task. 
For example, systems should allow the user to:
(1)  organize data into different ways, according to  the type of information or the level of detail she wishes to explore (e.g., \cite{bsps15}); 
(2) modify approximation criteria, thresholds, sampling rates, etc. (e.g., \cite{KamatJTN14});
(3)   define her own operations for data manipulation and analysis 
(e.g., aggregation, statistical, filtering functions), etc. 
Furthermore, systems should automatically adjust their parameters, by taking into account the  environment setting (e.g., screen resolution, memory size) \cite{JugelJM15,bsps15,JugelJHM14a}.

Beyond the personalization, modern systems should provide mechanisms that assist the user and reduce the effort needed on their part.
In this direction, several approaches have been recently developed.
In what follows, we mention some of the most common ones. 
Several systems   assist users by recommending visualization that seems to be more useful or capture  surprising and/or interesting data; e.g.,  \cite{WongsuphasawatM16,VartakMPP14,Key2012}. 
Other approaches help users to discover interest areas in the dataset; 
by capturing user interests, they guide her to interesting  data parts; e.g., \cite{DimitriadouPD14}. 
Finally, in other cases systems provide explanations regarding data trends and anomalies; e.g., \cite{WM13}.

%
%
%
%



\begin{table*}[!t]
\scriptsize
\centering
\caption{Generic Visualization Systems}
\label{tab:generic}
\setlength{\tabcolsep}{2.5pt}
\renewcommand{\arraystretch}{1.4}%
\begin{tabular}{l c  c c c c c c c c c c c }
\tline\vspace{-6pt}
\\
\textbf{System} &
\textbf{Year} &
\textbf{Data Types}$^\star$ &
\textbf{Vis.\ Types}$^{\star\star}$ &
\textbf{Recomm.\ } &
\textbf{Preferences} &
\textbf{Statistics}& 
\textbf{Sampling} &
\textbf{Aggregation}& 
\textbf{Incr.\ } &
\textbf{Disk} &
\textbf{Domain} &
\textbf{App.\ Type} \vspace{1pt}\\
\dlineB
%
%
%
\textbf{Rhizomer} \cite{BGG12} 	&	2006	&	 \textsf{N, T, S, H, G} 	&	 C, M, T, TL 	&	 \yes 	&	\no 	&	 \no 	&	 \no 	&	 \no 	&	 \no 	&	 \no 	&	 generic 	&	  Web\\
\rowcolor{gray!10}		
 \textbf{{VizBoard}} \cite{VoigtSGK12,vpm13,Polowinski13} 	&	2009	&	 \textsf{N, H} 	&	 C,S, T 	&	\yes 	&	 \yes 	&	 \no 	&	 \yes 	&	 \no 	&	\no 	&	 \no 	&	 generic 	&	  Web\\  
 \textbf{LODWheel} \cite{SDN11}	&	2011	&	    \textsf{N, S, G} 	&	C, G, M, P  	&	 \no  	&	 \no 	&	  \no 	&	 \no 	&	 \no 	&	 \no   	&	 \no 	&	 generic 	&	 Web\\
 \rowcolor{gray!10}		
 \textbf{SemLens} \cite{HeimLTE11}	&	2011	&	      \textsf{N} 	&	 S	&	  \no  	&	 \yes 	&	  \no 	&	 \no 	&	 \no 	&	 \no   	&	 \no 	&	 generic 	&	  Web\\
\textbf{LDVM} \cite{BrunettiAGKN13}  	&	2013	&	     \textsf{S, H, G} 	&	 B, M, T, TR  	&	 \yes 	&	 \no 	&	  \no 	&	 \no 	&	 \no 	&	  \no   	&	 \no 	&	 generic 	&	 Web\\
\rowcolor{gray!10}		
 \textbf{Payola} \cite{KHN13}  	&	2013	&	   \textsf{N, T, S, H, G} 	&	 C, CI, G, M,  T, TL, TR  	&	  \no  	&	 \no 	&	 \no  	&	 \no 	&	 \no 	&	 \no    	&	 \no 	&	 generic 	&	Web \\
\textbf{LDVizWiz} \cite{EURECOM+4380}    	&	2014	&	\textsf{S, H, G}  	&	M, P, TR 	&	  \yes 	&	 \no 	&	 \no  	&	 \no 	&	 \no 	&	  \no 	&	 \no 	&	 generic 	&	  Web\\
\rowcolor{gray!10}		
\textbf{SynopsViz}  \cite{bsp14, bsps15}	&	2014	&	 \textsf{N, T, H} 	&	  C, P, T, TL 	&	\yes  	&	 \yes  	&	 \yes 	&	 \no 	&	\yes	&	  \yes 	&	 \yes 	&	 generic 	&	 Web \\
 \textbf{Vis Wizard} \cite{TschinkelVMS14}	&	2014	&	    \textsf{N, T, S}	&	 B, C, M, P, PC, SG 	&	 \yes 	&	\yes 	&	  \no 	&	 \no 	&	 \no 	&	 \no   	&	 \no 	&	 generic 	&	  Web\\
 \rowcolor{gray!10}		
   \textbf{{LinkDaViz}} \cite{ThellmannGOS15}  	&	2015	&	  \textsf{N, T, S} 	&	 B, C, S, M, P 	&	 \yes 	&	 \yes 	&	 \no 	&	 \no 	&	 \no 	&	 \no 	&	 \no 	&	 generic 	&	  Web\\
\textbf{ViCoMap} \cite{RistoskiP15} 	&	2015	&	   \textsf{N, T, S} 	&	M 	&	\no 	&	 \no 	&	 \yes 	&	 \no 	&	 \no 	&	 \no  	&	 \no 	&	 generic 	&	 Web\\
 
\bline
\end{tabular}
{\vspace*{-6px}
\begin{flushleft} 
\hspace{0.5cm} 
 $^\star$
\textsf{N}: Numeric, 
\textsf{T}: Temporal,
\textsf{S}: Spatial,
\textsf{H}: Hierarchical (tree),
\textsf{G}: Graph (network)
\end{flushleft}}
\vspace{-15px}
{\begin{flushleft} 
\hspace{0.5cm} 
$^{\star\star}$  
B: \textit{bubble chart},  
C: \textit{chart},  
CI: \textit{circles}, 
G: \textit{graph}, 
M: \textit{map}, 
P: \textit{pie}, 
PC: \textit{parallel coordinates}, 
%
S: \textit{scatter}, 
SG: \textit{streamgraph}, 
T: \textit{treemap}, 
TL: \textit{timeline}, 
TR: \textit{tree}
\end{flushleft}}
\vspace{-21px}
\end{table*}

\section{Exploration \& Visualization\\ Systems}
 \label{sec:related}

This section reviews works related to exploration and  visualization  in the WoD.  
A large number of works studying issues related to WoD
visual exploration and analysis have been proposed in the literature \cite{DR11,MarieG14a,AlahmariTMW12}.
In what follows, we classify these works into the following categories: 
(1) Browsers and exploratory systems (Section~\ref{sec:browser}), 
(2) Generic visualization systems (Section~\ref{sec:generic}), 
(3) Domain, vocabulary \& device-specific visualization  systems (Section~\ref{sec:specific}), 
(4) Graph-based visualization systems (Section~\ref{sec:graph}), 
(5) Ontology visualization systems (Section~\ref{sec:onto}), 
and (6) Visualization libraries (Section~\ref{sec:libr}).

\subsection{Browsers \& Exploratory Systems}
\label{sec:browser}
\textit{WoD browsers} have been the first systems developed for WoD utilization and analysis \cite{DR11,AlahmariTMW12}.  
Similarly to the traditional ones, WoD browsers provide the functionality for link navigation and  
representation of WoD resources and their properties; thus enabling browsing and exploration of WoD in a most intuitive way. 
WoD browsers mainly use tabular views and links to provide navigation over the WoD resources.

\textit{Haystack} \cite{QuanK04}  is one of the first WoD browsers, it exploits \linebreak stylesheets 
in order to customize the data presentation.
Similarly, \textit{Disco}\footnote{\href{http://www4.wiwiss.fu-berlin.de/bizer/ng4j/disco}{www4.wiwiss.fu-berlin.de/bizer/ng4j/disco}}
  renders all information related to a 
particular RDF  resource as HTML table with property-value pairs.
\textit{Noadster} \cite{RutledgeOH05}   performs property-based data 
clustering in order to structure the results. 
\textit{Piggy Bank} \cite{HuynhMK05}  is a  Web browser plug-in,
 that allows users to convert  HTML  content into RDF.
\textit{LESS} \cite{AuerDD10} allows users to   create their own
 Web-based templates in order to aggregate and display WoD.
\textit{Tabulator} \cite{Berners-Lee2006}   another WoD browser, 
additionally provides maps and timeline visualizations.
 \textit{LENA} \cite{KochF08} provides different views of  data, following user's criteria that are expressed as SPARQL queries. 
 \textit{Visor} \cite{PopovSHS11} provides a  multi-pivot approach for exploring 
  graphs, allowing users to  explore  multiple nodes at a time, as well as to connect points of interest. 
Finally, in the context of faceted browsing,   $/$\textit{facet} \cite{HildebrandOH06},  \textit{Humboldt} \cite{KobilarovD08} and 
 \textit{gFacet}  \cite{HeimEZ10} provide faceted navigation over WoD resources.

\textit{Explorator} \cite{ASB09} is a WoD exploratory tool  that allows  users to 
browse a dataset by combining search and facets.
\textit{VisiNav} \cite{Harth10}  is a system that allows users to pose expressive exploratory-based queries. 
The system is built on top of following  concepts: keyword search, object focus, path traversal,
and facet selection.
\textit{Information Workbench} (IWB) \cite{HaaseSS11}
is a generic platform for semantic data management
offering several back-end (e.g., triple store) and front-end tools. 
Regarding the front-end, IWB offers a  flexible user interface for data exploration and visualization.
\textit{Marbles}\footnote{\href{http://mes.github.io/marbles/}{ mes.github.io/marbles}}
 formats  RDF triples using the Fresnel vocabulary  (a vocabulary for rendering RDF resources as HTML). 
Also, it retrieves information about a resource   by accessing Semantic Web indexes and search engines.
Finally, \textit{URI Burner}\footnote{\href{http://linkeddata.uriburner.com}{linkeddata.uriburner.com}} 
is a service which retrieves data about resources. For the requested resources, it generates an
RDF graph by exploiting existing ontologies and other knowledge from the Web.


\subsection{Generic Visualization Systems} 
\label{sec:generic}
 In the context of {WoD visual exploration}, 
there is a large number of  {generic visualization frameworks},
that offer a wide range of visualization types and operations. 
Next, we outline the best known systems in this category.

In Table~\ref{tab:generic} we provide an overview and compare several generic visualization systems.
The \textit{Year} column presents the released date.
The \textit{Data Types} column specifies the  supported data types. 
The \textit{Vis.\ Types} column presents  the types of  visualizations that are provided. 
The \textit{Recomm}.\ column indicates systems that offer recommendation mechanisms for visualization settings (e.g., appropriate visualization type, visualization  parameters).
The \textit{Preferences} column captures the ability of the users to 
apply data (e.g., filter, aggregate) or visual (e.g., increase abstraction) operations.
The \textit{Statistics} column captures the provision of statistics about the visualized data. 
The \textit{Sampling} column indicates systems that exploit 
techniques based on sampling and/or filtering.
The \textit{Aggregation} column indicates systems that exploit  techniques based on aggregation (e.g., binning, clustering).
The \textit{Incr}.\ column indicates systems that adopt incremental techniques;
i.e., the results/visualization are computed/generated  based on user interaction or as time progresses. 
Finally, the \textit{Disk} column indicates systems that use 
 external memory (e.g., file, database) to perform operations  during runtime (i.e., not just initially load data from disk).

\textit{Rhizomer} \cite{BGG12} provides WoD exploration based on a overview, zoom and filter workflow.  
Rhizomer  offers various types of visualizations such as maps, timelines, treemaps and charts.
\textit{VizBoard} \cite{VoigtSGK12,vpm13,Polowinski13} is an information
visualization workbench for WoD build on top of a mashup
platform. VizBoard presents datasets in a dashboard-like, composite,
and interactive visualization. Additionally, the system
provides visualization recommendations.
\textit{Payola} \cite{KHN13}  is a generic framework for WoD visualization and analysis. 
The framework offers a  variety of domain-specific (e.g., public procurement) analysis
plugins (i.e., analyzers), as well as several visualization techniques (e.g., graphs, tables).
In addition, Payola offers collaborative features for users to create and share analyzers. 
In Payola the visualizations can be customized according to ontologies used in the resulting data.

The  \textit{Linked Data Visualization Model} (LDVM) \cite{BrunettiAGKN13}
 provides an abstract visualization process for WoD datasets. 
 LDVM enables  the connection of different datasets with various kinds of visualizations in a dynamic way. 
The visualization process follows a four stage workflow:  Source data, 
Analytical abstraction, Visualization abstraction, and View. 
LDVM considers several visualization techniques, e.g., circle, sunburst, treemap, etc. 
Finally, the LDVM has been adopted in several use cases  \cite{KlimekHN15}.
\textit{Vis Wizard} \cite{TschinkelVMS14} is a Web-based visualization system, which exploits 
data semantics to simplify the process of setting up visualizations. 
Vis Wizard is able to analyse 
multiple datasets using brushing and linking methods.
Similarly, 
\textit{Linked Data Visualization Wizard} {(LDVizWiz)} \cite{EURECOM+4380} provides a 
semi-automatic way for the production of possible visualization for WoD datasets. 
In a same context, \textit{LinkDaViz} \cite{ThellmannGOS15} finds the suitable visualizations
for a give part of a dataset. The framework uses heuristic
data analysis and a visualization model in order to facilitate
automatic binding between data and visualization options.

\textit{Balloon Synopsis} \cite{SchlegelWSSGK14}  provides a  WoD visualizer based on 
HTML and JavaScript.  It adopts a node-centric visualization approach in a tile design.
Additionally, it supports automatic  information enhancement of the local RDF
data  by accessing either remote SPARQL endpoints or performing 
federated queries over endpoints using the Balloon Fusion service \cite{SchlegelSBGK14}.
Balloon Synopsis offers customizable filters,  namely    ontology templates,
for the users to handle and transform (e.g., filter, merge) input data. 
 \textit{LODWheel} \cite{SDN11}
is a Web-based visualizing tool which combines JavaScript libraries 
(e.g., MooWheel, JQPlot) in order to visualize RDF data in charts and graphs.
\textit{SemLens} \cite{HeimLTE11} is a visual tool that combines scatter plots and semantic lenses, offering 
visual discovery of correlations and patterns in data. 
Objects are arranged in  a scatter plot and are analysed using user-defined semantic lenses. 
\textit{ViCoMap} \cite{RistoskiP15} combines WoD
statistical analysis and visualization, in a Web-based tool, which offers correlation analysis and data visualization on maps.

Finally, \textit{SynopsViz}  \cite{bsp14, bsps15} is a Web-based visualization tool
built on top of a generic tree-based model. 
The adopted model performs a hierarchical aggregation, allowing efficient  personalized multilevel exploration over large datasets.
In order to provide scalability under different exploration scenarios, 
the model offers a method that incrementally constructs the hierarchy based on user's interaction, 
as well as a method that enables dynamic and efficient adaptation of the hierarchy to the user's preferences.
 \\
 
\begin{table*}[!t]
\scriptsize
\centering
\caption{Graph-based   Visualization Systems}
\label{tab:graph}
\setlength{\tabcolsep}{6.3pt}
\renewcommand{\arraystretch}{1.35}%
\begin{tabular}{l c  c c c c c c c c }
\tline\vspace{-6pt}
\\
\textbf{System} &
 \textbf{Year} &
 \textbf{Keyword} &
\textbf{Filter} &
\textbf{Sampling}& 
\textbf{Aggregation} &
 \textbf{Incr.\ } &
\textbf{Disk} &
\textbf{Domain} &
\textbf{App.\ Type} \vspace{1pt}\\
\dlineB

 \textbf{RDF-Gravity $^{\ref{fot:grav}}$}	&	2003	&	\yes	&	\yes	&	\no	&	\no	&	\no	&	\no	&	generic	&	Desktop	\\ \rowcolor{gray!10}		

\textbf{IsaViz} \cite{pietriga03} &	2003	&	\yes	&	\yes	&	\no	&	\no	&	\no &	\no	&	generic	&	Desktop	\\
\textbf{RDF graph visualizer} \cite{Sayers04} 	&	2004	&	\yes	&	\no	&	\no	&	\no	&	\no	&	\no	&	generic	&	Desktop	\\ \rowcolor{gray!10}		

\textbf{GrOWL}  \cite{KrivovWV07}	&	2007	&	\yes	&	\yes	&	\yes	&	\no	&	\no	&	\no	&	ontology	&	Desktop	\\
\textbf{NodeTrix} \cite{HenryFM07} 	&	2007	&	\no	&	\no	&	\no	&	\yes	&	\no	&	\no	&	ontology	&	Desktop	\\ \rowcolor{gray!10}		

\textbf{PGV}  \cite{DeligiannidisKS07} 	&	2007	&	\no	&	\no	&	\no	&	\no	&	\yes	&	\yes	&	generic	&	Desktop	\\
\textbf{Fenfire} \cite{HastrupCB08}  	&	2008	&	\no	&	\no	&	\no	&	\no	&	\no	&	\no	&	generic	&	Desktop	\\\rowcolor{gray!10}		

\textbf{Gephi} \cite{BastianHJ09}       	&	2009	&	\no	&	\yes	&	\yes	&	\yes	&	\no	&	\no	&	generic	&	Desktop	\\
\textbf{Trisolda} \cite{Trisolda09}	&	2010	&	\no	&	\no	&	\yes	&	\yes	&	\yes	&	\no	&	generic	&	Desktop	\\ \rowcolor{gray!10}		

\textbf{Cytospace}  \cite{SundaraAKDWCS10}	&	2010	&	\yes	&	\yes	&	\yes	&	\yes	&	\no	&	\yes	&	generic	&	Desktop	\\
\textbf{FlexViz} \cite{Falconer10}	&	2010	&	\yes	&	\yes	&	\no	&	\no	&	\no	&	\no	&	ontology	&	Web	\\ \rowcolor{gray!10}		

\textbf{RelFinder} \cite{HeimLS10} &	2010	&	\no	&	\no	&	\no	&	\no	&	\no	&	\no	&	generic	&	Web	\\
 \textbf{ZoomRDF} \cite{ZhangWTY10}	&	2010	&	\no	&	\no	&	\yes	&	\yes	&	\yes	&	\no	&	generic	&	Desktop	\\ \rowcolor{gray!10}		

\textbf{KC-Viz} \cite{MMP+11}	&	2011	&	\no	&	\no	&	\yes	&	\no	&	\no	&	\no	&	ontology	&	Desktop	\\
\textbf{LODWheel} \cite{SDN11} 	&	2011	&	\no	&	\yes	&	\no	&	\yes	&	\no	&	\no	&	generic	&	Web	\\ \rowcolor{gray!10}		

\textbf{GLOW}   \cite{HopRFH12}	&	2012	&	\no	&	\no	&	\yes	&	\yes	&	\no	&	\no	&	ontology	&	Desktop	\\
\textbf{Lodlive} \cite{CamardaMA12} 	&	2012	&	\yes	&	\no	&	\no	&	\no	&	\no	&	\no	&	generic	&	Web	\\ \rowcolor{gray!10}		

\textbf{OntoTrix} \cite{BachPL13} 	&	2013	&	\no	&	\no	&	\yes	&	\yes	&	\no	&	\no	&	ontology	&	Desktop	\\
\textbf{LODeX} \cite{BenedettiPB14} &	2014	&	\no	&	\no	&	\yes	&	\yes	&	\no	&	\no	&	generic	&	Web	\\ \rowcolor{gray!10}		

\textbf{VOWL 2} \cite{LNFT15,Lohmann14}	&	2014	&	\no	&	\no	&	\no	&	\no	&	\no	&	\no	&	ontology	&	Web	\\
\textbf{graphVizdb} \cite{BikakisLKG16,Bikakis15} 	&	2015	&	\yes	&	\yes	&	\yes	&	\no	&	\no	&	\yes	&	generic	&	Web	\\
 
\bline
\end{tabular}
\end{table*}


\vspace{-2mm}
\subsection{Domain, Vocabulary \& Device-specific \\ Visualization Systems}
\label{sec:specific}
In this section, we present systems that target visualization needs for
specific types of data and domains, RDF vocabularies or  {devices}. 

Several systems 
focus on visualizing and exploring geo-spatial data.
\textit{Map4rdf} \cite{Map4rdf} is a faceted browsing tool that enables
RDF datasets to be visualized on an OSM or Google Map. 
\textit{Facete}  \cite{StadlerMA14}  is an exploration and visualization tool
for SPARQL accessible data, offering faceted  filtering functionalities. 
\textit{SexTant} \cite{BeretaNKKK13} and \textit{Spacetime}  \cite{Valsecchi14} focus on visualizing and  exploring  time-evolving geo-spatial data.  
The \textit{LinkedGeoData Browser}  \cite{StadlerLHA12} is a faceted
browser and editor which is developed in the context of 
LinkedGeoData  project.
Finally, in the same context \textit{DBpedia Atlas} \cite{ValsecchiABTM15} offers exploration over the DBpedia dataset by exploiting the dataset's spatial data. 
Furthermore,  in the context of linked university data,   \textit{VISUalization
Playground} (VISU) \cite{AKSH13} is an interactive tool for specifying and creating
visualizations using the contents of linked university data cloud. 
Particularly, VISU offers a  novel SPARQL interface for creating
data visualizations.
Query results from selected SPARQL endpoints are visualized with Google Charts.
 
A variety of systems target multidimensional WoD modelled with the Data Cube vocabulary. 
\textit{CubeViz} \cite{ErmilovMLA13,SMB+12}  is a faceted browser  for exploring statistical data.
The tool provides data visualizations using different types of charts (i.e., line, bar, column, area and pie).
The \textit{Payola Data Cube Vocabulary} \cite{HelmichKN14}
adopts the LDVM stages \cite{BrunettiAGKN13}  in order to visualize RDF data described  by the Data Cube vocabulary. 
The same types of charts as in CubeViz are provided in this tool. 
%
The \textit{OpenCube  Toolkit} \cite{Kalampokis14} offers several tools related to statistical WoD.
For example, \textit{OpenCube Browser} explores RDF data cubes by
presenting a two-dimensional table. 
Additionally,  the \textit{OpenCube Map View} offers
interactive map-based visualizations of RDF data cubes based on their geo-spatial dimension.
The \textit{Linked Data Cubes Explorer} (LDCE) \cite{KampgenH14}
allows users to explore and analyse statistical datasets.
Finally, \cite{petrou2014e} offers several map and chart visualizations of 
demographic, social and statistical linked cube data.


%
Regarding device-specific systems, \textit{DBpedia Mobile} \cite{BeckerB09}    is a location-aware mobile application
for exploring and visualizing DBpedia resources. 
\textit{Who's Who} \cite{CanoDH11}  is an  application  for exploring and visualizing information
focusing on several issues that appear in the mobile environment. 
For example, the application considers the
usability and data processing challenges related to 
the small display size and   limited resources of the mobile devices.

\subsection{Graph-based Visualization Systems}
\label{sec:graph}
A large number of systems visualize WoD
datasets adopting a \textit{graph-based} (a.k.a., node-link) approach \cite{MazumdarPELC15}.
In Table~\ref{tab:graph} we provide an overview and compare several graph-based visualization systems.
 Table~\ref{tab:graph} is structured in a similar way to Table~\ref{tab:generic}.
Additionally, in this table the \textit{Keyword} column indicates systems that provide keyword search functionality.
The \textit{Filter} column indicates systems that provide mechanisms for data filtering. 
Note that, Table~\ref{tab:graph} also includes the ontology visualization systems (Section~\ref{sec:onto})  that follow a  node-link approach (indicated by using the term "ontology" in the Domain column).

 \textit{RelFinder} \cite{HeimLS10} is a Web-based tool that offers  interactive discovery and visualization 
of relationships (i.e., connections) between selected WoD resources. 
\textit{Fenfire} \cite{HastrupCB08} and    \textit{Lodlive} \cite{CamardaMA12}
are exploratory tools that allow users to browse WoD using interactive graphs.
Starting from a given URI, the user can explore WoD by following the links.
\textit{LODeX} \cite{BenedettiPB14}
is a tool that generates a representative summary of a WoD source. 
The tool takes as input a SPARQL endpoint and generates a visual (graph-based) summary of the WoD source, 
accompanied  by statistical and structural information of the source. 
\textit{IsaViz} \cite{pietriga03} allows users to zoom and navigate over the RDF graph, 
and also it offers several "edit" operations (e.g., delete/add/rename nodes and edges).
In the same context, \textit{graphVizdb} \cite{BikakisLKG16,Bikakis15}  is built on top of spatial and
database techniques offering interactive visualization over very large (RDF) graphs. 
\textit{ZoomRDF} \cite{ZhangWTY10} employs a space-optimized visualization algorithm
 in order to increase the number of resources which are   displayed. 
\textit{Trisolda} \cite{Trisolda09}  proposes a hierarchical RDF graph visualization. 
It adopts clustering techniques in order to merge graph nodes. 
%
\textit{Paged Graph Visualization} (PGV) \cite{DeligiannidisKS07} utilizes  a Ferris-Wheel approach to display nodes with high degree.
\textit{RDF graph visualizer} \cite{Sayers04} adopts a node-centric approach to visualize RDF graphs.
Rather than trying to visualize the  whole graph, nodes of interest (i.e., staring nodes) are discovered
by searching over nodes labels; then the user can interactively navigate over the graph.
\textit{RDF-Gravity}\footnote{\label{fot:grav}\href{http://semweb.salzburgresearch.at/apps/rdf-gravity}{semweb.salzburgresearch.at/apps/rdf-gravity}} visualizes RDF and OWL data. 
 It offers filtering, keyword search and editing the   graph layout. 
 Also, the nodes can be displayed in different colors and  shapes based on their RDF types.
A different approach has been adopted in \cite{SundaraAKDWCS10},
where  sampling techniques have been exploited. 
Finally, \textit{Gephi} \cite{BastianHJ09} is a generic  tool that
offers several visualization and analysis features over graph data.

\subsection{Ontology Visualization Systems}
\label{sec:onto}
The problems of \textit{ontology visualization and exploration} 
have been extensively studied in several research areas (e.g., biology, chemistry).  
In what follows we focus  on graph-based ontology visualization systems that have been  developed 
in the WoD context \cite{FuN14,DudasZS14,HaagLNE14,LanzenbergerSR09,KatiforiHLVG07}.
In most systems, ontologies are visualized following the node-link paradigm \cite{LNFT15,Lohmann14,HopRFH12,MMP+11,Boinski10,Falconer10,HLR14,LiebigN05,alani03,KrivovWV07,StoreyNMBFE02} 
\footnote{\href{http://protegewiki.stanford.edu/wiki/OntoGraf}{protegewiki.stanford.edu/wiki/OntoGraf}}$^{,}$\footnote{\href{http://protegewiki.stanford.edu/wiki/OWLViz}{protegewiki.stanford.edu/wiki/OWLViz}}.
On the other hand, \textit{CropCircles} \cite{WangP06a} uses a geometric containment approach, representing the class hierarchy as a set of concentric circles.
Furthermore,   hybrids approaches are adopted in other works. 
\textit{Knoocks} \cite{KriglsteinM08} combines containment-based 
and node-link approaches. 
In this work, ontologies are visualized as 
nested blocks where each block is depicted as a
rectangle containing a sub-branch shown as tree map.
Finally, \textit{OntoTrix} \cite{BachPL13} and  \textit{NodeTrix} \cite{HenryFM07} 
use  node-link and adjacency matrix representations.



%

\subsection{Visualization Libraries}
\label{sec:libr}
Finally, there is a variety of Javascript libraries which allow WoD visualizations to be embedded in Web pages. 
\textit{Sgvizler} \cite{S12}  is a JavaScript wrapper  for visualizing  SPARQL results. 
Sgvizler allows users to  specify SPARQL Select queries directly into HTML elements.
Sgvizler uses   Google Charts to generate the output,
offering numerous visualizations types such as charts, treemaps, graphs, timelines,  etc.
\textit{Visualbox} \cite{GA13}  provides an environment where users can 
build and debug SPARQL queries in order to retrieve WoD; 
then, a set of visualization templates is provided to  visualize results. 
Visualbox  uses several visualization libraries like Google Charts and 
D3 \cite{BostockOH11}, offering 14 visualization types.
%


  \section{Discussion}
  \label{sec:disc}
  
 In this section we discuss to which extent the systems developed in the 
 WoD context fulfilled the nowadays requirements, 
 focussing on  performance and scalability issues, availability of personalized services facilities for assisting users through exploration.
 
 As previously mentioned, most of WoD exploration and visualization systems 
 do not handle issues related to performance and scalability. 
 They basically adopt traditional techniques in order to handle small sets of data. 
 
 As we can observe  from Table~\ref{tab:generic},   generic systems
  support several types of data (e.g., numeric, temporal, graph, spatial) and provide a plethora of visualization types. Additionally, an increasing number of recent systems 
(e.g.,  LinkDaViz, Vis Wizard, LDVizWiz, LDVM) focus on  providing recommendation mechanisms.  
Particularity, these systems mainly recommend the most 
 suitable visualization technique  by considering the type of input data. 
 
 Regarding visual scalability, as we can see in  Table~\ref{tab:generic}, 
 none of the systems, with the exceptions of SynopsViz and VizBoard cases, adopt 
 approximation techniques (i.e., sampling/filtering, aggregation). 
 Hence, the existing approaches assume that all the examined  data objects 
 can be presented on the screen and handled by traditional visualization techniques. 
 Due to this assumption, the current systems restrict their applicability to 
 small sets of data.  
 
In conjunction with the limited visual scalability, most of the 
existing systems (except for SynopsViz) do not exploit external memory during runtime.
Particularly, they initially load all the examined objects in main memory, assuming that the main memory is large enough.
An alternative approach is adopted by the  SynopsViz system, which 
incrementally retrieves data and generates visualizations based on user interaction. 
As a result, each time, only a part of the examined dataset needs to be loaded in main memory.

The graph-based exploration and visualization systems are presented in Table~\ref{tab:graph}.
These systems are   of great importance in WoD, due to the graph structure of the RDF data model. 
Although several  systems offer sampling or aggregation mechanisms, 
most of these systems load the whole graph in main memory. 
Given the large memory requirements of graph layout algorithms 
in order to draw a large graph, 
the current WoD systems are restricted to handle small sized graphs.

In order to be able to handle large graphs, modern WoD systems should adopt 
more sophisticated techniques  similar to those proposed by the information visualization community. 
Particularly,   state-of-the-art systems for exploring large graphs 
utilize  hierarchical aggregation approaches where the graph
is recursively decomposed into smaller sub-graphs (in most cases using clustering and partitioning) that form a hierarchy of abstraction layers \cite{LBW15,Archambault2011,LinCTWKC13,ArchambaultMA08,ArchambaultMA07,AbelloHK06,ZinsmaierBDS12,Auber04,BastianHJ09,RodriguesTPTTF13,TominskiAS09}.
Other approaches adopt edge bundling techniques
 which  aggregate graph edges to bundles  \cite{Gansner2011,Ersoy2011,Phan2005,Lambert2010,Cui2008,Holten2006}.
Beyond hierarchical approaches, WoD systems should also consider 
disk-based implementations, such as \cite{Bikakis15,AbelloHK06,RodriguesTTFL06,SundaraAKDWCS10,TominskiAS09}.

To sum up, WoD community should consider scalability and performance
as vital requirements for the development  of the future exploration and visualization systems. 
Handing large datasets is crucial in the Big Data era. 
Therefore,    in what follows we summarize some possible 
directions for the future WoD exploration and visualization systems.
Approximation techniques such as sampling and aggregation that 
have been widely used in systems from database and information visualization communities,
have to be adopted and adjusted to WoD data and requirements.  
Systems should be integrated with disk structures,  retrieving data dynamically during runtime. 
Also caching and prefetching techniques may be exploited; e.g., \cite{TauheedHSMA12,KalininCZ14,JayachandranTKN14,bcs15,ChanXGH08,KhanSA14,DoshiRW03}. 
Data structures and indexes should be developed focusing on  WoD tasks and data,
such as Nanocubes \cite{LinsKS13} in the context of  spatio-temporal  data exploration, and HETree \cite{bsps15}   in  numeric and temporal datasets. 
Finally, considering users' perspective, beyond visualization   recommendations, modern WoD systems should provide more sophisticated mechanisms that  capture users' preferences and assist them throughout large data exploration and analysis tasks. 
  

%

\begin{thebibliography}{100}

\fontsize{9pt}{9.3pt}\selectfont

\bibitem{AbelloHK06}
J.~Abello, F.~van Ham, and N.~Krishnan.
\newblock {ASK}-{G}raph{V}iew: {A} {L}arge {S}cale {G}raph {V}isualization
  {S}ystem.
\newblock {\em TVCG}, 12(5), 2006.

\bibitem{AgarwalMPMMS13}
S.~Agarwal, B.~Mozafari, A.~Panda, H.~Milner, S.~Madden, and I.~Stoica.
\newblock {B}link{DB}: {Q}ueries with {B}ounded {E}rrors and {B}ounded
  {R}esponse {T}imes on {V}ery {L}arge {D}ata.
\newblock In {\em EuroSys}, 2013.

\bibitem{Ahlberg96}
C.~Ahlberg.
\newblock {S}potfire: {A}n {I}nformation {E}xploration {E}nvironment.
\newblock {\em {SIGMOD} Record}, 25(4), 1996.

\bibitem{AlahmariTMW12}
F.~Alahmari, J.~A. Thom, L.~Magee, and W.~Wong.
\newblock {E}valuating {S}emantic {B}rowsers for {C}onsuming {L}inked {D}ata.
\newblock In {\em ADC}, 2012.

\bibitem{alani03}
H.~Alani.
\newblock {TGV}iz{T}ab: {A}n {O}ntology {V}isualisation {E}xtension for
  {P}rotege.
\newblock In {\em Workshop on Visualizing Information in Knowledge
  Engineering}, 2003.

\bibitem{AKSH13}
M.~Alonen, T.~Kauppinen, O.~Suominen, and E.~Hyv{\"{o}}nen.
\newblock {E}xploring the {L}inked {U}niversity {D}ata with {V}isualization
  {T}ools.
\newblock In {\em ESWC}, 2013.

\bibitem{ASB09}
S.~F.~C. Ara\'ujo, D.~Schwabe, and S.~D.~J. Barbosa.
\newblock {E}xperimenting with {E}xplorator: a {D}irect {M}anipulation
  {G}eneric {RDF} {B}rowser and {Q}uerying {T}ool.
\newblock In {\em Visual Interfaces to the Social and the Semantic Web}, 2009.

\bibitem{ArchambaultMA07}
D.~Archambault, T.~Munzner, and D.~Auber.
\newblock {G}rouse: {F}eature-{B}ased, {S}teerable {G}raph {H}ierarchy
  {E}xploration.
\newblock In {\em EuroVis}, 2007.

\bibitem{ArchambaultMA08}
D.~Archambault, T.~Munzner, and D.~Auber.
\newblock {G}rouse{F}locks: {S}teerable {E}xploration of {G}raph {H}ierarchy
  {S}pace.
\newblock {\em TVCG}, 14(4), 2008.

\bibitem{Archambault2011}
D.~Archambault, T.~Munzner, and D.~Auber.
\newblock {T}ugging {G}raphs {F}aster: {E}fficiently {M}odifying
  {P}ath-{P}reserving {H}ierarchies for {B}rowsing {P}aths.
\newblock {\em TVCG}, 17(3), 2011.

\bibitem{EURECOM+4380}
G.~A. {A}temezing and R.~{T}roncy.
\newblock {T}owards a linked-data based visualization wizard.
\newblock In {\em COLD}, 2014.

\bibitem{Auber04}
D.~Auber.
\newblock {T}ulip - {A} {H}uge {G}raph {V}isualization {F}ramework.
\newblock In {\em Graph Drawing Software}. 2004.

\bibitem{AuerDD10}
S.~Auer, R.~Doehring, and S.~Dietzold.
\newblock {LESS} -- {T}emplate-{B}ased {S}yndication and {P}resentation of
  {L}inked {D}ata.
\newblock In {\em ESWC}, 2010.

\bibitem{BachPL13}
B.~Bach, E.~Pietriga, and I.~Liccardi.
\newblock {V}isualizing {P}opulated {O}ntologies with {O}nto{T}rix.
\newblock {\em IJSWIS}, 9(4), 2013.

\bibitem{BastianHJ09}
M.~Bastian, S.~Heymann, and M.~Jacomy.
\newblock {G}ephi: {A}n {O}pen {S}ource {S}oftware for {E}xploring and
  {M}anipulating {N}etworks.
\newblock In {\em ICWSM}, 2009.

\bibitem{bcs15}
L.~Battle, R.~Chang, and M.~Stonebraker.
\newblock {D}ynamic {P}refetching of {D}ata {T}iles for {I}nteractive
  {V}isualization, 2015.
\newblock Technical Report.

\bibitem{BattleSC13}
L.~Battle, M.~Stonebraker, and R.~Chang.
\newblock {D}ynamic reduction of query result sets for interactive
  visualizaton.
\newblock In {\em BigData}, 2013.

\bibitem{BeckerB09}
C.~Becker and C.~Bizer.
\newblock {E}xploring the {G}eospatial {S}emantic {W}eb with {DB}pedia
  {M}obile.
\newblock {\em J. Web Sem.}, 7(4), 2009.

\bibitem{BenedettiPB14}
F.~Benedetti, L.~Po, and S.~Bergamaschi.
\newblock {A} {V}isual {S}ummary for {L}inked {O}pen {D}ata sources.
\newblock In {\em ISWC}, 2014.

\bibitem{BeretaNKKK13}
K.~Bereta, C.~Nikolaou, M.~Karpathiotakis, K.~Kyzirakos, and M.~Koubarakis.
\newblock {S}ex{T}ant: {V}isualizing {T}ime-{E}volving {L}inked {G}eospatial
  {D}ata.
\newblock In {\em ISWC}, 2013.

\bibitem{Berners-Lee2006}
T.~Berners-Lee, Y.~Chen, L.~Chilton, D.~Connolly, R.~Dhanaraj, J.~Hollenbach,
  A.~Lerer, and D.~Sheets.
\newblock {T}abulator: {E}xploring and {A}nalyzing linked data on the
  {S}emantic {W}eb.
\newblock In {\em SWUI}, 2006.

\bibitem{Bikakis15}
N.~Bikakis, J.~Liagouris, M.~Krommyda, G.~Papastefanatos, and T.~Sellis.
\newblock {T}owards {S}calable {V}isual {E}xploration of {V}ery {L}arge {RDF}
  {G}raphs.
\newblock In {\em ESWC}, 2015.

\bibitem{BikakisLKG16}
N.~Bikakis, J.~Liagouris, M.~Krommyda, G.~Papastefanatos, and T.~Sellis.
\newblock graph{V}izdb: {A} {S}calable {P}latform for {I}nteractive {L}arge
  {G}raph {V}isualization.
\newblock In {\em {ICDE}}, 2016.

\bibitem{bp16}
N.~Bikakis and G.~Papastefanatos.
\newblock {V}isual {E}xploration and {A}nalytics of {B}ig {D}ata: {C}hallenges
  and {A}pproaches, 2016.

\bibitem{bsps15}
N.~Bikakis, G.~Papastefanatos, M.~Skourla, and T.~Sellis.
\newblock {A} {H}ierarchical {A}ggregation {F}ramework for {E}fficient
  {M}ultilevel {V}isual {E}xploration and {A}nalysis, 2015.
\newblock Techn. Rep., \url{http://arxiv.org/abs/1511.04750}.

\bibitem{bsp14}
N.~Bikakis, M.~Skourla, and G.~Papastefanatos.
\newblock rdf:{S}ynops{V}iz - {A} {F}ramework for {H}ierarchical {L}inked
  {D}ata {V}isual {E}xploration and {A}nalysis.
\newblock In {\em ESWC}, 2014.

\bibitem{Boinski10}
T.~Boinski, A.~Jaworska, R.~Kleczkowski, and P.~Kunowski.
\newblock {O}ntology visualization.
\newblock In {\em ITNGSWUI}, 2010.

\bibitem{BostockOH11}
M.~Bostock, V.~Ogievetsky, and J.~Heer.
\newblock {D}{\({^3}\)} {D}ata-{D}riven {D}ocuments.
\newblock {\em TVCG}, 17(12), 2011.

\bibitem{BrunettiAGKN13}
J.~M. Brunetti, S.~Auer, R.~Garc{\'{\i}}a, J.~Kl{\'{\i}}mek, and
  M.~Necask{\'{y}}.
\newblock {F}ormal {L}inked {D}ata {V}isualization {M}odel.
\newblock In {\em iiWAS}, 2013.

\bibitem{BGG12}
J.~M. Brunetti, R.~Gil, and R.~Garc{\'{\i}}a.
\newblock {F}acets and {P}ivoting for {F}lexible and {U}sable {L}inked {D}ata
  {E}xploration.
\newblock In {\em Interacting with Linked Data Workshop}, 2012.

\bibitem{CamardaMA12}
D.~V. Camarda, S.~Mazzini, and A.~Antonuccio.
\newblock {L}od{L}ive, exploring the web of data.
\newblock In {\em I-SEMANTICS}, 2012.

\bibitem{CanoDH11}
A.~E. Cano, A.~Dadzie, and M.~Hartmann.
\newblock \emph{Who's Who} - {A} {L}inked {D}ata {V}isualisation {T}ool for
  {M}obile {E}nvironments.
\newblock In {\em ESWC}, 2011.

\bibitem{ChanXGH08}
S.~Chan, L.~Xiao, J.~Gerth, and P.~Hanrahan.
\newblock {M}aintaining interactivity while exploring massive time series.
\newblock In {\em IEEE VAST}, 2008.

\bibitem{Cui2008}
W.~Cui, H.~Zhou, H.~Qu, P.~C. Wong, and X.~Li.
\newblock {G}eometry-{B}ased {E}dge {C}lustering for {G}raph {V}isualization.
\newblock {\em TVCG}, 14(6), 2008.

\bibitem{DR11}
A.~Dadzie and M.~Rowe.
\newblock {A}pproaches to visualising {L}inked {D}ata: {A} survey.
\newblock {\em Semantic Web}, 2(2), 2011.

\bibitem{DeligiannidisKS07}
L.~Deligiannidis, K.~Kochut, and A.~P. Sheth.
\newblock {RDF} data exploration and visualization.
\newblock In {\em Workshop on CyberInfrastructure: Information Management in
  eScience}, 2007.

\bibitem{DimitriadouPD14}
K.~Dimitriadou, O.~Papaemmanouil, and Y.~Diao.
\newblock {E}xplore-by-{E}xample: {A}n {A}utomatic {Q}uery {S}teering
  {F}ramework for {I}nteractive {D}ata {E}xploration.
\newblock In {\em {SIGMOD}}, 2014.

\bibitem{Trisolda09}
J.~Dokulil and J.~Katreniakov\'{a}.
\newblock {U}sing {C}lusters in {RDF} {V}isualization.
\newblock In {\em Advances in Semantic Processing}, 2009.

\bibitem{DoshiRW03}
P.~R. Doshi, E.~A. Rundensteiner, and M.~O. Ward.
\newblock {P}refetching for {V}isual {D}ata {E}xploration.
\newblock In {\em DASFAA}, 2003.

\bibitem{DudasZS14}
M.~Dud{\'{a}}s, O.~Zamazal, and V.~Sv{\'{a}}tek.
\newblock {R}oadmapping and {N}avigating in the {O}ntology {V}isualization
  {L}andscape.
\newblock In {\em EKAW}, 2014.

\bibitem{EMJ16}
A.~Eldawy, M.~Mokbel, and C.~Jonathan.
\newblock {H}adoop{V}iz: {A} {M}ap{R}educe {F}ramework for {E}xtensible
  {V}isualization of {B}ig {S}patial {D}ata.
\newblock In {\em {ICDE}}, 2016.

\bibitem{EF10}
N.~Elmqvist and J.~Fekete.
\newblock {H}ierarchical {A}ggregation for {I}nformation {V}isualization:
  {O}verview, {T}echniques, and {D}esign {G}uidelines.
\newblock {\em TVCG}, 16(3), 2010.

\bibitem{ErmilovMLA13}
I.~Ermilov, M.~Martin, J.~Lehmann, and S.~Auer.
\newblock {L}inked {O}pen {D}ata {S}tatistics: {C}ollection and {E}xploitation.
\newblock In {\em Knowledge Engineering and the Semantic Web}, 2013.

\bibitem{Ersoy2011}
O.~Ersoy, C.~Hurter, F.~V. Paulovich, G.~Cantareiro, and A.~Telea.
\newblock {S}keleton-{B}ased {E}dge {B}undling for {G}raph {V}isualization.
\newblock {\em TVCG}, 17(12), 2011.

\bibitem{Falconer10}
S.~Falconer, C.~Callendar, and M.-A. Storey.
\newblock {A} {V}isualization {S}ervice for the {S}emantic {W}eb.
\newblock In {\em Knowledge Engineering and Management by the Masses}. 2010.

\bibitem{FisherPDs12}
D.~Fisher, I.~O. Popov, S.~M. Drucker, and M.~C. Schraefel.
\newblock {T}rust {M}e, {I}'m {P}artially {R}ight: {I}ncremental
  {V}isualization {L}ets {A}nalysts {E}xplore {L}arge {D}atasets {F}aster.
\newblock In {\em CHI}, 2012.

\bibitem{FuN14}
B.~Fu, N.~F. Noy, and M.-A. Storey.
\newblock {E}ye {T}racking the {U}ser {E}xperience - {A}n {E}valuation of
  {O}ntology {V}isualization {T}echniques.
\newblock {\em Semantic Web Journal}, 2015.

\bibitem{Gansner2011}
E.~R. Gansner, Y.~Hu, S.~C. North, and C.~E. Scheidegger.
\newblock {M}ultilevel {A}gglomerative {E}dge {B}undling for {V}isualizing
  {L}arge {G}raphs.
\newblock In {\em PacificVis}, 2011.

\bibitem{GGL15}
P.~Godfrey, J.~Gryz, and P.~Lasek.
\newblock {I}nteractive {V}isualization of {L}arge {D}ata {S}ets, 2015.
\newblock Technical Report.

\bibitem{GA13}
A.~Graves.
\newblock {C}reation of {V}isualizations {B}ased on {L}inked {D}ata.
\newblock In {\em WIMS}, 2013.

\bibitem{HaagLNE14}
F.~Haag, S.~Lohmann, S.~Negru, and T.~Ertl.
\newblock {O}nto{V}i{B}e: {A}n {O}ntology {V}isualization {B}enchmark.
\newblock In {\em VISUAL}, 2014.

\bibitem{HaaseSS11}
P.~Haase, M.~Schmidt, and A.~Schwarte.
\newblock {T}he {I}nformation {W}orkbench as a {S}elf-{S}ervice {P}latform for
  {L}inked {D}ata {A}pplications.
\newblock In {\em COLD}, 2011.

\bibitem{Harth10}
A.~Harth.
\newblock {V}isi{N}av: {A} system for visual search and navigation on web data.
\newblock {\em J. Web Sem.}, 8(4), 2010.

\bibitem{HastrupCB08}
T.~Hastrup, R.~Cyganiak, and U.~Bojars.
\newblock {B}rowsing {L}inked {D}ata with {F}enfire.
\newblock In {\em WWW}, 2008.

\bibitem{HeerK12b}
J.~Heer and S.~Kandel.
\newblock {I}nteractive {A}nalysis of {B}ig {D}ata.
\newblock {\em {ACM} Crossroads}, 19(1), 2012.

\bibitem{HeerS12}
J.~Heer and B.~Shneiderman.
\newblock {I}nteractive {D}ynamics for {V}isual {A}nalysis.
\newblock {\em Commun. {ACM}}, 55(4), 2012.

\bibitem{HeimEZ10}
P.~Heim, T.~Ertl, and J.~Ziegler.
\newblock {F}acet {G}raphs: {C}omplex {S}emantic {Q}uerying {M}ade {E}asy.
\newblock In {\em ESWC}, 2010.

\bibitem{HeimLS10}
P.~Heim, S.~Lohmann, and T.~Stegemann.
\newblock {I}nteractive {R}elationship {D}iscovery via the {S}emantic {W}eb.
\newblock In {\em ESWC}, 2010.

\bibitem{HeimLTE11}
P.~Heim, S.~Lohmann, D.~Tsendragchaa, and T.~Ertl.
\newblock {S}em{L}ens: visual analysis of semantic data with scatter plots and
  semantic lenses.
\newblock In {\em I-SEMANTICS}, 2011.

\bibitem{HelmichKN14}
J.~Helmich, J.~Kl{\'{\i}}mek, and M.~Necask{\'{y}}.
\newblock {V}isualizing {RDF} {D}ata {C}ubes {U}sing the {L}inked {D}ata
  {V}isualization {M}odel.
\newblock In {\em ESWC}, 2014.

\bibitem{HenryFM07}
N.~Henry, J.~Fekete, and M.~J. McGuffin.
\newblock {N}ode{T}rix: a {H}ybrid {V}isualization of {S}ocial {N}etworks.
\newblock {\em TVCG}, 13(6), 2007.

\bibitem{HildebrandOH06}
M.~Hildebrand, J.~van Ossenbruggen, and L.~Hardman.
\newblock /facet: {A} {B}rowser for {H}eterogeneous {S}emantic {W}eb
  {R}epositories.
\newblock In {\em ISWC}, 2006.

\bibitem{Holten2006}
D.~Holten.
\newblock {H}ierarchical {E}dge {B}undles: {V}isualization of {A}djacency
  {R}elations in {H}ierarchical {D}ata.
\newblock {\em TVCG}, 12(5), 2006.

\bibitem{HopRFH12}
W.~Hop, S.~de~Ridder, F.~Frasincar, and F.~Hogenboom.
\newblock {U}sing {H}ierarchical {E}dge {B}undles to visualize complex
  ontologies in {GLOW}.
\newblock In {\em ACM SAC}, 2012.

\bibitem{HLR14}
A.~Hussain, K.~Latif, A.~Rextin, A.~Hayat, and M.~Alam.
\newblock {S}calable {V}isualization of {S}emantic {N}ets using {P}ower-{L}aw
  {G}raphs.
\newblock {\em AMIS}, 8(1), 2014.

\bibitem{HuynhMK05}
D.~Huynh, S.~Mazzocchi, and D.~R. Karger.
\newblock {P}iggy {B}ank: {E}xperience the {S}emantic {W}eb {I}nside {Y}our
  {W}eb {B}rowser.
\newblock In {\em ISWC}, 2005.

\bibitem{IdreosKM07}
S.~Idreos, M.~L. Kersten, and S.~Manegold.
\newblock {D}atabase {C}racking.
\newblock In {\em CIDR}, 2007.

\bibitem{IdreosPC15}
S.~Idreos, O.~Papaemmanouil, and S.~Chaudhuri.
\newblock {O}verview of {D}ata {E}xploration {T}echniques.
\newblock In {\em {SIGMOD}}, 2015.

\bibitem{ImVM13}
J.~Im, F.~G. Villegas, and M.~J. McGuffin.
\newblock {V}is{R}educe: {F}ast and {R}esponsive {I}ncremental {I}nformation
  {V}isualization of {L}arge {D}atasets.
\newblock In {\em BigData}, 2013.

\bibitem{JayachandranTKN14}
P.~Jayachandran, K.~Tunga, N.~Kamat, and A.~Nandi.
\newblock {C}ombining {U}ser {I}nteraction, {S}peculative {Q}uery {E}xecution
  and {S}ampling in the {DICE} {S}ystem.
\newblock {\em PVLDB}, 7(13), 2014.

\bibitem{RodriguesTPTTF13}
J.~F.~R. Jr., H.~Tong, J.~Pan, A.~J.~M. Traina, C.~T. Jr., and C.~Faloutsos.
\newblock {L}arge {G}raph {A}nalysis in the {GM}ine {S}ystem.
\newblock {\em TKDE}, 25(1), 2013.

\bibitem{RodriguesTTFL06}
J.~F.~R. Jr., H.~Tong, A.~J.~M. Traina, C.~Faloutsos, and J.~Leskovec.
\newblock {GM}ine: {A} {S}ystem for {S}calable, {I}nteractive {G}raph
  {V}isualization and {M}ining.
\newblock In {\em VLDB}, 2006.

\bibitem{JugelJHM14a}
U.~Jugel, Z.~Jerzak, G.~Hackenbroich, and V.~Markl.
\newblock {F}aster {V}isual {A}nalytics through {P}ixel-{P}erfect
  {A}ggregation.
\newblock {\em PVLDB}, 7(13), 2014.

\bibitem{JugelJM15}
U.~Jugel, Z.~Jerzak, G.~Hackenbroich, and V.~Markl.
\newblock {VDDA}: automatic visualization-driven data aggregation in relational
  databases.
\newblock {\em VLDBJ}, 2015.

\bibitem{Kalampokis14}
E.~Kalampokis, A.~Nikolov, P.~Haase, R.~Cyganiak, A.~Stasiewicz, A.~Karamanou,
  M.~Zotou, D.~Zeginis, E.~Tambouris, and K.~A. Tarabanis.
\newblock {E}xploiting {L}inked {D}ata {C}ubes with {O}pen{C}ube {T}oolkit.
\newblock In {\em ISWC}, 2014.

\bibitem{KalininCZ14}
A.~Kalinin, U.~{\c{C}}etintemel, and S.~B. Zdonik.
\newblock {I}nteractive {D}ata {E}xploration {U}sing {S}emantic {W}indows.
\newblock In {\em {SIGMOD}}, 2014.

\bibitem{KalininCZ15}
A.~Kalinin, U.~{\c{C}}etintemel, and S.~B. Zdonik.
\newblock {S}earchlight: {E}nabling {I}ntegrated {S}earch and {E}xploration
  over {L}arge {M}ultidimensional {D}ata.
\newblock {\em PVLDB}, 8(10), 2015.

\bibitem{KamatJTN14}
N.~Kamat, P.~Jayachandran, K.~Tunga, and A.~Nandi.
\newblock {D}istributed and {I}nteractive {C}ube {E}xploration.
\newblock In {\em {ICDE}}, 2014.

\bibitem{KampgenH14}
B.~K{\"{a}}mpgen and A.~Harth.
\newblock {OLAP4LD} - {A} {F}ramework for {B}uilding {A}nalysis {A}pplications
  {O}ver {G}overnmental {S}tatistics.
\newblock In {\em ESWC}, 2014.

\bibitem{KatiforiHLVG07}
A.~Katifori, C.~Halatsis, G.~Lepouras, C.~Vassilakis, and E.~G. Giannopoulou.
\newblock {O}ntology visualization methods - a survey.
\newblock {\em {ACM} Comput. Surv.}, 39(4), 2007.

\bibitem{KeimK94}
D.~Keim and H.-P. Kriegel.
\newblock {V}is{DB}: {D}atabase {E}xploration using {M}ultidimensional
  {V}isualization.
\newblock {\em {IEEE} Computer Graphics and Applications}, 14(5), 1994.

\bibitem{Key2012}
A.~Key, B.~Howe, D.~Perry, and C.~R. Aragon.
\newblock {V}iz{D}eck: {S}elf-organizing {D}ashboards for {V}isual {A}nalytics.
\newblock In {\em {SIGMOD}}, 2012.

\bibitem{KhanSA14}
H.~A. Khan, M.~A. Sharaf, and A.~Albarrak.
\newblock {D}iv{IDE}: efficient diversification for interactive data
  exploration.
\newblock In {\em SSDBM}, 2014.

\bibitem{KHN13}
J.~Kl{\'{\i}}mek, J.~Helmich, and M.~Necask{\'{y}}.
\newblock {P}ayola: {C}ollaborative {L}inked {D}ata {A}nalysis and
  {V}isualization {F}ramework.
\newblock In {\em ESWC}, 2013.

\bibitem{KlimekHN15}
J.~Kl{\'{\i}}mek, J.~Helmich, and M.~Necask{\'{y}}.
\newblock {U}se {C}ases for {L}inked {D}ata {V}isualization {M}odel.
\newblock In {\em LDOW}, 2015.

\bibitem{KobilarovD08}
G.~Kobilarov and I.~Dickinson.
\newblock {H}umboldt: {E}xploring {L}inked {D}ata.
\newblock In {\em LDOW}, 2008.

\bibitem{KochF08}
J.~Koch and T.~Franz.
\newblock {LENA} - {B}rowsing {RDF} {D}ata {M}ore {C}omplex {T}han {F}oaf.
\newblock In {\em ISWC}, 2008.

\bibitem{KriglsteinM08}
S.~Kriglstein and R.~Motschnig{-}Pitrik.
\newblock {K}noocks: {N}ew {V}isualization {A}pproach for {O}ntologies.
\newblock In {\em Conference on Information Visualisation}, 2008.

\bibitem{KrivovWV07}
S.~Krivov, R.~Williams, and F.~Villa.
\newblock {G}r{OWL}: {A} tool for visualization and editing of {OWL}
  ontologies.
\newblock {\em J. Web Sem.}, 5(2), 2007.

\bibitem{Lambert2010}
A.~Lambert, R.~Bourqui, and D.~Auber.
\newblock {W}inding {R}oads: {R}outing {E}dges into {B}undles.
\newblock {\em CGF}, 29(3), 2010.

\bibitem{LanzenbergerSR09}
M.~Lanzenberger, J.~Sampson, and M.~Rester.
\newblock {V}isualization in {O}ntology {T}ools.
\newblock In {\em CISIS}, 2009.

\bibitem{Map4rdf}
A.~d. Leon, F.~Wisniewki, B.~Villaz\'{o}n-Terrazas, and O.~Corcho.
\newblock {M}ap4rdf- {F}aceted {B}rowser for {G}eospatial {D}atasets.
\newblock In {\em Using Open Data: policy modeling, citizen empowerment, data
  journalism}, 2012.

\bibitem{LBW15}
C.~Li, G.~Baciu, and Y.~Wang.
\newblock {M}odul{G}raph: {M}odularity-based {V}isualization of {M}assive
  {G}raphs.
\newblock In {\em Visualization in High Performance Computing}, 2015.

\bibitem{LiebigN05}
T.~Liebig and O.~Noppens.
\newblock {O}nto{T}rack: {A} semantic approach for ontology authoring.
\newblock {\em J. Web Sem.}, 3(2-3), 2005.

\bibitem{LinCTWKC13}
Z.~Lin, N.~Cao, H.~Tong, F.~Wang, U.~Kang, and D.~H.~P. Chau.
\newblock {D}emonstrating {I}nteractive {M}ulti-resolution {L}arge {G}raph
  {E}xploration.
\newblock In {\em ICDM Workshops}, 2013.

\bibitem{LinsKS13}
L.~D. Lins, J.~T. Klosowski, and C.~E. Scheidegger.
\newblock {N}anocubes for {R}eal-{T}ime {E}xploration of {S}patiotemporal
  {D}atasets.
\newblock {\em TVCG}, 19(12), 2013.

\bibitem{LiuJH13}
Z.~Liu, B.~Jiang, and J.~Heer.
\newblock \emph{imMens}: {R}eal-time {V}isual {Q}uerying of {B}ig {D}ata.
\newblock {\em CGF}, 32(3):421--430, 2013.

\bibitem{LivnyRBCDLMW97a}
M.~Livny, R.~Ramakrishnan, K.~S. Beyer, G.~Chen, D.~Donjerkovic, S.~Lawande,
  J.~Myllymaki, and R.~K. Wenger.
\newblock {DEV}ise: {I}ntegrated {Q}uerying and {V}isual {E}xploration of
  {L}arge {D}atasets.
\newblock In {\em {SIGMOD}}, 1997.

\bibitem{Lohmann14}
S.~Lohmann, S.~Negru, F.~Haag, and T.~Ertl.
\newblock {VOWL} 2: {U}ser-{O}riented {V}isualization of {O}ntologies.
\newblock In {\em EKAW}, 2014.

\bibitem{LNFT15}
S.~Lohmann, S.~Negru, F.~Haag, and T.~Ertl.
\newblock {V}isualizing {O}ntologies with {VOWL}.
\newblock {\em Semantic Web Journal}, 2015.

\bibitem{MarieG14a}
N.~Marie and F.~L. Gandon.
\newblock {S}urvey of {L}inked {D}ata {B}ased {E}xploration {S}ystems.
\newblock In {\em IESD}, 2014.

\bibitem{MazumdarPELC15}
S.~Mazumdar, D.~Petrelli, K.~Elbedweihy, V.~Lanfranchi, and F.~Ciravegna.
\newblock {A}ffective graphs: {T}he visual appeal of {L}inked {D}ata.
\newblock {\em Semantic Web}, 6(3), 2015.

\bibitem{MortonBGM14}
K.~Morton, M.~Balazinska, D.~Grossman, and J.~D. Mackinlay.
\newblock {S}upport the {D}ata {E}nthusiast: {C}hallenges for
  {N}ext-{G}eneration {D}ata-{A}nalysis {S}ystems.
\newblock {\em PVLDB}, 7(6), 2014.

\bibitem{MMP+11}
E.~Motta, P.~Mulholland, S.~Peroni, M.~d'Aquin, J.~M.
  G{\'{o}}mez{-}P{\'{e}}rez, V.~Mendez, and F.~Zablith.
\newblock {A} {N}ovel {A}pproach to {V}isualizing and {N}avigating
  {O}ntologies.
\newblock In {\em ISWC}, 2011.

\bibitem{ParkCM15}
Y.~Park, M.~J. Cafarella, and B.~Mozafari.
\newblock {V}isualization-{A}ware {S}ampling for {V}ery {L}arge {D}atabases.
\newblock In {\em {ICDE}}, 2016.

\bibitem{petrou2014e}
I.~Petrou, M.~Meimaris, and G.~Papastefanatos.
\newblock {T}owards a methodology for publishing {L}inked {O}pen {S}tatistical
  {D}ata.
\newblock {\em eJournal of eDemocracy \& Open Government}, 6(1), 2014.

\bibitem{Phan2005}
D.~Phan, L.~Xiao, R.~B. Yeh, P.~Hanrahan, and T.~Winograd.
\newblock {F}low {M}ap {L}ayout.
\newblock In {\em InfoVis}, 2005.

\bibitem{pietriga03}
E.~Pietriga.
\newblock {I}sa{V}iz: a {V}isual {E}nvironment for {B}rowsing and {A}uthoring
  {RDF} {M}odels.
\newblock In {\em WWW}, 2002.

\bibitem{Polowinski13}
J.~Polowinski and M.~Voigt.
\newblock {VISO}: {A} {S}hared, {F}ormal {K}nowledge {B}ase {A}s a {F}oundation
  for {S}emi-automatic {I}nfovis {S}ystems.
\newblock In {\em {CHI}}, 2013.

\bibitem{PopovSHS11}
I.~O. Popov, M.~M.~C. Schraefel, W.~Hall, and N.~Shadbolt.
\newblock {C}onnecting the {D}ots: {A} {M}ulti-pivot {A}pproach to {D}ata
  {E}xploration.
\newblock In {\em ISWC}, 2011.

\bibitem{QuanK04}
D.~A. Quan and R.~Karger.
\newblock {H}ow to make a semantic web browser.
\newblock In {\em WWW}, 2004.

\bibitem{RistoskiP15}
P.~Ristoski and H.~Paulheim.
\newblock {V}isual {A}nalysis of {S}tatistical {D}ata on {M}aps using {L}inked
  {O}pen {D}ata.
\newblock In {\em ESWC}, 2015.

\bibitem{RutledgeOH05}
L.~Rutledge, J.~van Ossenbruggen, and L.~Hardman.
\newblock {M}aking {RDF} presentable: integrated global and local semantic
  {W}eb browsing.
\newblock In {\em WWW}, 2005.

\bibitem{SMB+12}
P.~E.~R. Salas, F.~M.~D. Mota, K.~K. Breitman, M.~A. Casanova, M.~Martin, and
  S.~Auer.
\newblock {P}ublishing {S}tatistical {D}ata on the {W}eb.
\newblock {\em Int. J. Semantic Computing}, 6(4), 2012.

\bibitem{Sayers04}
C.~Sayers.
\newblock {N}ode-centric {RDF} {G}raph {V}isualization, 2004.
\newblock Technical Report HP Laboratories.

\bibitem{SchlegelSBGK14}
K.~Schlegel, F.~Stegmaier, S.~Bayerl, M.~Granitzer, and H.~Kosch.
\newblock {B}alloon {F}usion: {SPARQL} {R}ewriting based on {U}nified
  {C}o-{R}eference {I}nformation.
\newblock In {\em DESWeb}, 2014.

\bibitem{SchlegelWSSGK14}
K.~Schlegel, T.~Wei{\ss}gerber, F.~Stegmaier, C.~Seifert, M.~Granitzer, and
  H.~Kosch.
\newblock {B}alloon {S}ynopsis: {A} {M}odern {N}ode-{C}entric {RDF} {V}iewer
  and {B}rowser for the {W}eb.
\newblock In {\em ESWC}, 2014.

\bibitem{Shneiderman96}
B.~Shneiderman.
\newblock {T}he {E}yes {H}ave {I}t: {A} {T}ask by {D}ata {T}ype {T}axonomy for
  {I}nformation {V}isualizations.
\newblock In {\em IEEE Symposium on Visual Languages}, 1996.

\bibitem{Shneiderman08}
B.~Shneiderman.
\newblock {E}xtreme {V}isualization: {S}queezing a {B}illion {R}ecords into a
  {M}illion {P}ixels.
\newblock In {\em {SIGMOD}}, 2008.

\bibitem{S12}
M.~G. Skj{\ae}veland.
\newblock {S}gvizler: {A} {J}ava{S}cript {W}rapper for {E}asy {V}isualization
  of {SPARQL} {R}esult {S}ets.
\newblock In {\em ESWC}, 2012.

\bibitem{StadlerLHA12}
C.~Stadler, J.~Lehmann, K.~H{\"{o}}ffner, and S.~Auer.
\newblock {L}inked{G}eo{D}ata: {A} core for a web of spatial open data.
\newblock {\em Semantic Web}, 3(4), 2012.

\bibitem{StadlerMA14}
C.~Stadler, M.~Martin, and S.~Auer.
\newblock {E}xploring the web of spatial data with facete.
\newblock In {\em WWW}, 2014.

\bibitem{StolperPG14}
C.~D. Stolper, A.~Perer, and D.~Gotz.
\newblock {P}rogressive {V}isual {A}nalytics: {U}ser-{D}riven {V}isual
  {E}xploration of {I}n-{P}rogress {A}nalytics.
\newblock {\em TVCG}, 20(12), 2014.

\bibitem{StolteH00}
C.~Stolte and P.~Hanrahan.
\newblock {P}olaris: {A} {S}ystem for {Q}uery, {A}nalysis and {V}isualization
  of {M}ulti-{D}imensional {R}elational {D}atabases.
\newblock In {\em InfoVis}, 2000.

\bibitem{StoreyNMBFE02}
M.~D. Storey, N.~F. Noy, M.~A. Musen, C.~Best, R.~W. Fergerson, and N.~A.
  Ernst.
\newblock {J}ambalaya: an interactive environment for exploring ontologies.
\newblock In {\em {IUI}}, 2002.

\bibitem{SDN11}
M.~Stuhr, D.~Roman, and D.~Norheim.
\newblock {LODW}heel - {J}ava{S}cript-based {V}isualization of {RDF} {D}ata.
\newblock In {\em COLD}, 2011.

\bibitem{SundaraAKDWCS10}
S.~Sundara, M.~Atre, V.~Kolovski, S.~Das, Z.~Wu, E.~I. Chong, and
  J.~Srinivasan.
\newblock {V}isualizing large-scale {RDF} data using {S}ubsets, {S}ummaries,
  and {S}ampling in {O}racle.
\newblock In {\em {ICDE}}, 2010.

\bibitem{TauheedHSMA12}
F.~Tauheed, T.~Heinis, F.~Sch{\"{u}}rmann, H.~Markram, and A.~Ailamaki.
\newblock {SCOUT:} {P}refetching for {L}atent {F}eature {F}ollowing {Q}ueries.
\newblock {\em PVLDB}, 5(11), 2012.

\bibitem{ThellmannGOS15}
K.~Thellmann, M.~Galkin, F.~Orlandi, and S.~Auer.
\newblock {L}ink{D}a{V}iz - {A}utomatic {B}inding of {L}inked {D}ata to
  {V}isualizations.
\newblock In {\em ISWC}, 2015.

\bibitem{TominskiAS09}
C.~Tominski, J.~Abello, and H.~Schumann.
\newblock {CGV} - {A}n {I}nteractive {G}raph {V}isualization {S}ystem.
\newblock {\em Computers {\&} Graphics}, 33(6), 2009.

\bibitem{TschinkelVMS14}
G.~Tschinkel, E.~E. Veas, B.~Mutlu, and V.~Sabol.
\newblock {U}sing {S}emantics for {I}nteractive {V}isual {A}nalysis of {L}inked
  {O}pen {D}ata.
\newblock In {\em ISWC}, 2014.

\bibitem{ValsecchiABTM15}
F.~Valsecchi, M.~Abrate, C.~Bacciu, M.~Tesconi, and A.~Marchetti.
\newblock {DB}pedia {A}tlas: {M}apping the {U}ncharted {L}ands of {L}inked
  {D}ata.
\newblock In {\em LDOW}, 2015.

\bibitem{Valsecchi14}
F.~Valsecchi and M.~Ronchetti.
\newblock {S}pacetime: a {T}wo {D}imensions {S}earch and {V}isualisation
  {E}ngine {B}ased on {L}inked{D}ata.
\newblock In {\em SEMAPRO}, 2014.

\bibitem{VartakMPP14}
M.~Vartak, S.~Madden, A.~G. Parameswaran, and N.~Polyzotis.
\newblock {SEEDB:} {A}utomatically {G}enerating {Q}uery {V}isualizations.
\newblock {\em PVLDB}, 7(13), 2014.

\bibitem{VoigtSGK12}
M.~Voigt, S.~Pietschmann, L.~Grammel, and K.~Mei{\ss}ner.
\newblock {C}ontext-aware {R}ecommendation of {V}isualization {C}omponents.
\newblock In {\em eKNOW}, 2012.

\bibitem{vpm13}
M.~Voigt, S.~Pietschmann, and K.~Mei{\ss}ner.
\newblock {A} {S}emantics-{B}ased, {E}nd-{U}ser-{C}entered {I}nformation
  {V}isualization {P}rocess for {S}emantic {W}eb {D}ata.
\newblock In {\em Semantic Models for Adaptive Interactive Systems}. 2013.

\bibitem{WangP06a}
T.~D. Wang and B.~Parsia.
\newblock {C}rop{C}ircles: {T}opology {S}ensitive {V}isualization of {OWL}
  {C}lass {H}ierarchies.
\newblock In {\em ISWC}, 2006.

\bibitem{hw13}
H.~Wickham.
\newblock {B}in-{S}ummarise-{S}mooth: {A} {F}ramework for {V}isualising {L}arge
  {D}ata.
\newblock Technical report, 2013.

\bibitem{WongsuphasawatM16}
K.~Wongsuphasawat, D.~Moritz, A.~Anand, J.~D. Mackinlay, B.~Howe, and J.~Heer.
\newblock {V}oyager: {E}xploratory {A}nalysis via {F}aceted {B}rowsing of
  {V}isualization {R}ecommendations.
\newblock {\em TVCG}, 22(1), 2016.

\bibitem{WuBM14}
E.~Wu, L.~Battle, and S.~R. Madden.
\newblock {T}he {C}ase for {D}ata {V}isualization {M}anagement {S}ystems.
\newblock {\em PVLDB}, 7(10), 2014.

\bibitem{WM13}
E.~Wu and S.~Madden.
\newblock {S}corpion: {E}xplaining {A}way {O}utliers in {A}ggregate {Q}ueries.
\newblock {\em PVLDB}, 6(8), 2013.

\bibitem{ZhangWTY10}
K.~Zhang, H.~Wang, D.~T. Tran, and Y.~Yu.
\newblock {Z}oom{RDF}: semantic fisheye zooming on {RDF} data.
\newblock In {\em WWW}, 2010.

\bibitem{ZinsmaierBDS12}
M.~Zinsmaier, U.~Brandes, O.~Deussen, and H.~Strobelt.
\newblock {I}nteractive {L}evel-of-{D}etail {R}endering of {L}arge {G}raphs.
\newblock {\em TVCG}, 18(12), 2012.

\bibitem{ZoumpatianosIP14}
K.~Zoumpatianos, S.~Idreos, and T.~Palpanas.
\newblock {I}ndexing for interactive exploration of big data series.
\newblock In {\em {SIGMOD}}, 2014.

\end{thebibliography}

\end{document}